\documentclass{article}
\usepackage[margin=1in]{geometry}
\usepackage{soul}
\usepackage{url}
\usepackage[hidelinks]{hyperref}
\usepackage[utf8]{inputenc}
\usepackage{caption}
\usepackage{graphicx}
\usepackage{amsmath}
\usepackage{amssymb}
\usepackage{booktabs}
\usepackage{subfig}
\usepackage{float}
\usepackage{bm}
\usepackage{natbib}
\newtheorem{prop}{Proposition}

\newcommand{\squishlist}{
 \begin{list}{$\bullet$}
  { \setlength{\itemsep}{0pt}
     \setlength{\parsep}{2pt}
     \setlength{\topsep}{2pt}
     \setlength{\partopsep}{0pt}
     \setlength{\leftmargin}{1em}
     \setlength{\labelwidth}{1em}
     \setlength{\labelsep}{0.5em} } }

     \newcommand{\squishend}{
  \end{list}  }

\title{TuSeRACT: Turn-Sample-Based Real-Time Traffic Signal Control}
\author{Srishti Dhamija and Pradeep Varakantham \\ School of Information Systems, Singapore Management University}
\date{}

\begin{document}
\maketitle

\begin{abstract}
	Real-time traffic signal control is a challenging problem owing to constantly changing traffic demand patterns, limited planning time and various sources of uncertainty (e.g., turn movements, vehicle detection) in the real world. SURTRAC (Scalable URban TRAffic Control) is a recently developed traffic signal control approach which computes delay-minimizing and coordinated (across neighbouring traffic lights) schedules of oncoming vehicle clusters in real time. To ensure real-time responsiveness in the presence of turn-induced uncertainty, SURTRAC computes schedules whic­h minimize the delay for the expected turn movements as opposed to minimizing the expected delay under turn-induced uncertainty. This approximation ensures real-time tractability, but degrades solution quality in the presence of turn-induced uncertainty. To address this limitation, we introduce TuSeRACT (Turn Sample based Real-time trAffic signal ConTrol), a distributed sample-based scheduling approach to traffic signal control. Unlike SURTRAC, TuSeRACT computes schedules that minimize expected delay over sampled turn movements of observed traffic, and communicates samples of traffic outflows to neighbouring intersections. We formulate this sample-based scheduling problem as a constraint program and empirically evaluate our approach on synthetic traffic networks. Our approach provides substantially lower mean vehicular waiting times relative to SURTRAC. 	
\end{abstract}

\section{Introduction}
Sub-optimal traffic signal control contributes significantly to urban traffic congestion \citep{chin2004temporary} due to: (i) poor allocation of green light time to competing traffic streams; (ii) lack of coordination between traffic lights; (iii) inability to respond in real-time to changing traffic patterns. Improvements to traffic signal mechanisms can reduce traffic congestion and therefore increase the effective capacity of existing road networks. However, real-time traffic signal control is challenging because scalable, network-wide optimal control must be achieved under limited planning time, constantly changing demand patterns and real-world uncertainty \citep{xie2014coping,cai2009performance,yu2006stochastic}. While centralized traffic-responsive signal control systems \citep{sims1980sydney,robertson1991optimizing,luyanda2003acs} can reduce network-wide delays, they are not scalable due to exponentially increasing problem sizes with network expansion. Recent research \citep{sen1997controlled,gartner2002optimized,bazzan2005distributed,kuyer2008multiagent} has therefore focused on decentralized, coordinated adaptive traffic signal control mechanisms and  SURTRAC (Scalable Urban Traffic Control) \citep{xie2012coordination,xie2012schedule} is the leading approach in this category that has been deployed on real traffic control systems in Pittsburgh.  

SURTRAC  is a real-time, distributed, schedule-driven approach, where each intersection independently schedules clusters of incoming vehicles along different directions (referred to as phases) and communicates projected expected outgoing traffic clusters to traffic signals at neighbouring intersections. SURTRAC has been widely deployed due to its scalability and real-time responsiveness and has provided significant improvements over existing systems.

Despite its scalability and effectiveness, SURTRAC has two fundamental limitations: (a) To ensure real-time response given the vehicular turn movement uncertainties, instead of minimizing expected  delay, SURTRAC minimizes delay for the expected scenario of vehicular turn movements. (b) In order to ensure scalable coordination among traffic signals, expected outgoing traffic is communicated to traffic signals at neighbouring intersections. These approximations ensure real-time tractability but are particularly limiting in the presence of turn-induced uncertainty.

To address these key limitations, we introduce TuSeRACT (Turn-Sample-based Real-time trAffic signal ConTrol), a turn sample-driven distributed scheduling approach to traffic signal control. Unlike SURTRAC, TuSeRACT: (i) optimizes expected delay over a set of turn samples; and (ii) communicates samples of projected outgoing traffic to traffic signals at neighbouring intersections.

This is achieved through a combination of  three key contributions. First, we provide a Constraint Programming (CP) formulation that is based on Sample Average Approximation (SAA) \citep{kleywegt2002sample} and employs turn samples for controlling each individual traffic signal. Second, we provide a novel communication mechanism between traffic signals at neighbouring intersections that is based on samples of outgoing traffic. Third, we reuse the strategy generated previously to guide search for better strategy. 

On multiple benchmark problems considered in the literature, we demonstrate that TuSeRACT is able to significantly outperform the leading approach for real-time traffic signal control, SURTRAC. We were able to consistently reduce the mean vehicular delay (by up to 50\%) across multiple traffic networks under varying traffic demand conditions.

\section{Distributed Traffic Signal Control}
\label{sec:model}

Traffic signal control systems aim to give right-of-way (green time) to competing streams of oncoming traffic at intersections in a manner that minimizes network-wide vehicular waiting time. Real time traffic signal control typically distributes this decision process to individual intersections so as to ensure real-time responsiveness to changing trafffic patterns. We provide a theoretical justification for such a distribution of control to individual intersections (with traffic inflow information from neighbouring intersections) under some assumptions. 

For each intersection $i \in V$ in the network, the traffic signal control problem at a decision step is defined as the tuple $\left<\lambda_i, \delta_{-i}\right>$, where $\lambda_i$ refers to the local context for intersection $i$ (operating constraints, observed traffic, initial conditions) and $\delta_{-i}$ refers to the non-local context (traffic conditions) of relevance to $i$ at that decision step which is communicated from one or more intersections in $V-\{i\}$. Since we are interested in real-time control, all the traffic of relevance at an decision step is coming from immediate neighbouring intersections of $i$ i.e. $N_i$.

\begin{figure}
	\centering
	\subfloat[An intersection with multiple entry and exit approaches. Arrows indicate the direction of traffic flow.]{\includegraphics[scale=0.25]{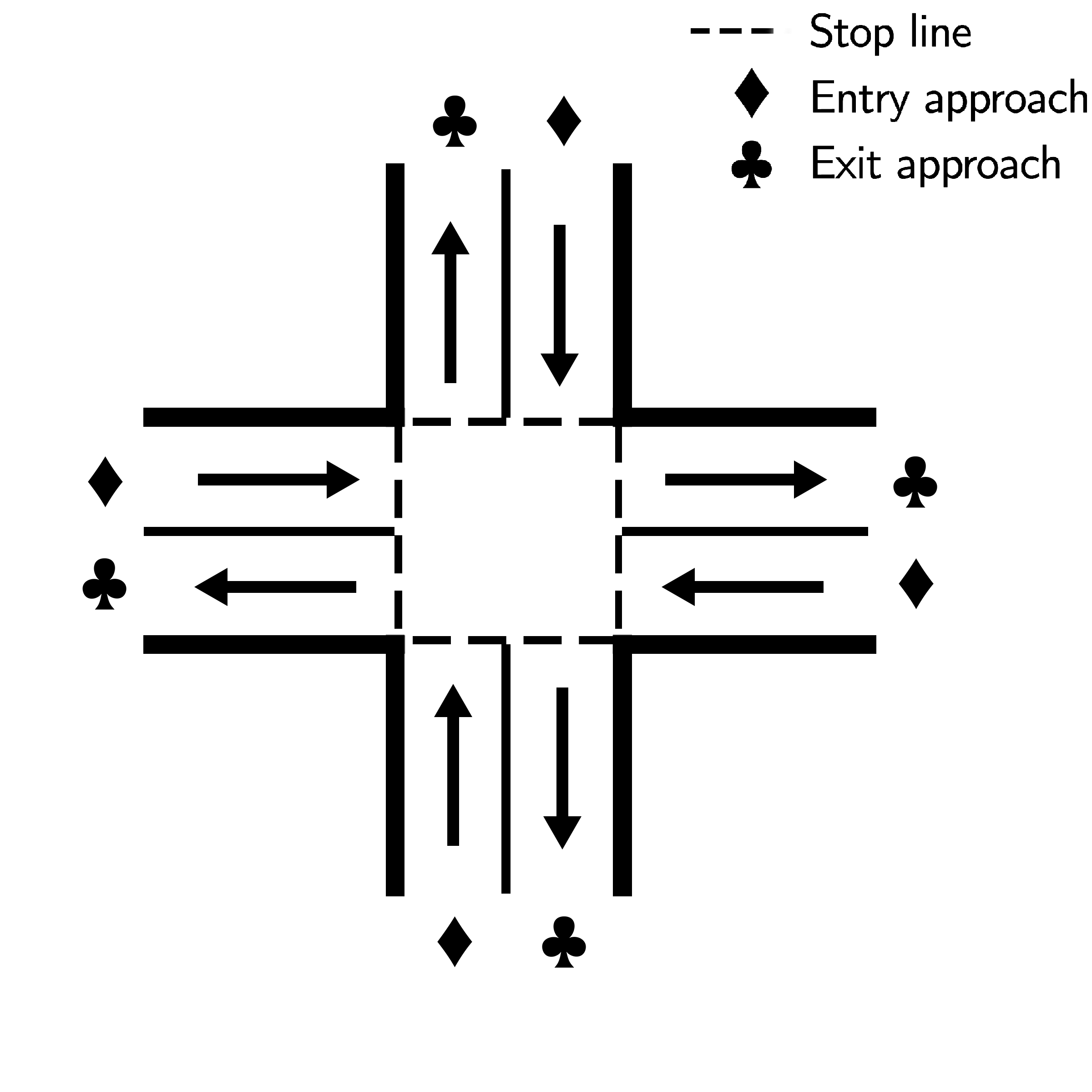}
	\label{fig:intersection}}

	\subfloat[Phase design for a 4-phase intersection. Each arrow represents a turn movement, i.e. traffic flow from an entry road to an exit road at the intersection. Turn movements which can safely be given right-of-way simultaneously are grouped into a \textit{phase}.]{\includegraphics[scale=0.5]{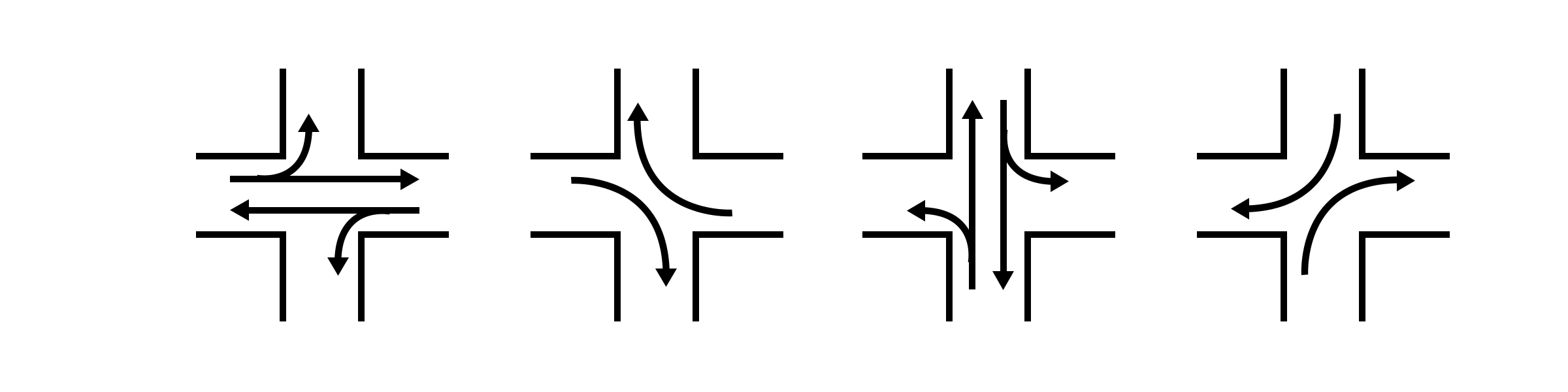}
	\label{fig:phases}}
\end{figure}

The local context $\lambda_i$ is defined as the tuple: 
{\begin{align}
\left<\Gamma_{i}, \{p_{i}^{\tau}\}_{\tau \in \Gamma}, \phi_i, \delta_i, \theta^0_i\right> \nonumber
\end{align}}
\squishlist
    \item[$\Gamma_i$:] is the set of permissible turn movements at  intersection $i$. Each turn movement, $\tau \in \Gamma_i$ is defined as $\left<e,f\right>$ and represents a traffic movement from entry road $e$ to exit road $f$ of the intersection.
    \item[$p_i^{\tau}$:] represents the probability for turn movement $\tau$ at  intersection $i$. In practice, turn probabilities are estimated as moving averages using recently observed turn movements.
	\item[$\phi_i$:] is the {\em phase model} for the intersection and is defined as an ordered set of phases, i.e.,  $\left<\psi_0, \dots, \psi_{|\phi_i|-1}\right>$.  Traffic signals are usually required to cyclically give right-of-way to phases in $\phi_i$, that is $\psi_{(k+1) \bmod |\phi_i|}$ is given right-of-way after $\psi^k$. Each phase, $\psi_{k}$ gives right-of-way to a set of non-conflicting turn movements, as shown in Figure~\ref{fig:phases}. 
	Each phase $\psi^k$ is represented using the tuple: $$\left< {\cal T}^k, G_{min}^k, G^k_{max}, Y^k\right>$$
	\squishlist
		\item[${\cal T}^k$:] is the set of turns which have right-of-way during the phase $\psi^k$, where ${\cal T}^k \subset \Gamma_i$ and $\bigcup_{k=0}^{|\phi_i|-1} {\cal T}^k = \Gamma_i$
		\item[$G_{min}^k$ and $G^k_{max}$:] are the lower and upper bounds on the time for which $\psi^k$ has right-of-way (green time).
		\item[$Y^k$:] is the fixed inter-green time (informally yellow time) which must be applied after $\psi^k$, during which no phase has right-of-way to ensure safety. 
\squishend
	    $\phi_i$ thus specifies the phase order, turns permitted in each phase, minimum and maximum bounds on phase duration and inter-green time between phases at intersection $i$. 
	
	\item[$\delta_i$:] is the traffic detected on the intersection's entry roads (shown in Figure~\ref{fig:intersection}).
	\item[$\theta^0_i$:] represents the initial conditions at intersection $i$ and is given by the tuple $\left<\psi^0, g\right>$, where $\psi^0$ is the phase which has right-of-way (current phase) and $g$ is the time for which it has been green (current phase duration). The initial conditions determine the extension feasibility of the current phase.
\squishend

\noindent Given a planning horizon and $\left<\lambda_i, \delta_{-i}\right>$ at a decision point, we compute a {\em traffic signal timing plan}, $\pi_i$ for intersection $i$. A traffic signal timing plan specifies the duration of each phase, $\psi^k$ in every cycle $r$ and is represented as $\pi_{i}^{k,r}$. We compute a traffic signal timing plan that minimizes the expected delay for the vehicles approaching the intersection:
{\begin{align}
\min_{\pi_i} \mathbb{E}_{p_i, p_{N_i} } \Big[{\cal L}\Big( \delta_i^0, \ldots, \delta_{i}^{|\phi_i| - 1}\hspace{0.05in} | \hspace{0.05in} \delta_i, \lambda_i, \delta_{N_i}, \pi_i \Big) \Big] \label{eqn:tuseract}
\end{align}}
where $\delta_{i}^k$ represents the traffic on phase $k$ and ${\cal L}(.)$ represents the latency (or delay) random variable for each phase, $\psi^k$.  $p_i$ and $p_{N_i}$ determine the number of vehicles in each phase and consequently determine the delay/latency along each phase.

The computed traffic signal timing plan is then implemented up to the next decision point that can be before the planning horizon $H$.  The extent of commitment to formulated plans varies and is dependent on how quickly decisions have to be made. We consider systems which simply make an \textit{termination} or \textit{extension}  decision for the current phase $\psi^k$ and then recompute the plan to account for new traffic:
\begin{itemize}
	\item {\textit Termination}: The current phase $\psi^k$ is terminated and the next phase $\psi_{l}$ is given right-of-way for a minimum green time $G^{min}_{{k}}$ after an intergreen time $Y_{k}$. 
	\item {\textit Extension}: The current phase $\psi^k$ is extended for a duration $\epsilon > 0$.
\end{itemize}

\subsection{Dec-MDP Representation}
\begin{figure}
\centering\includegraphics[scale=0.3]{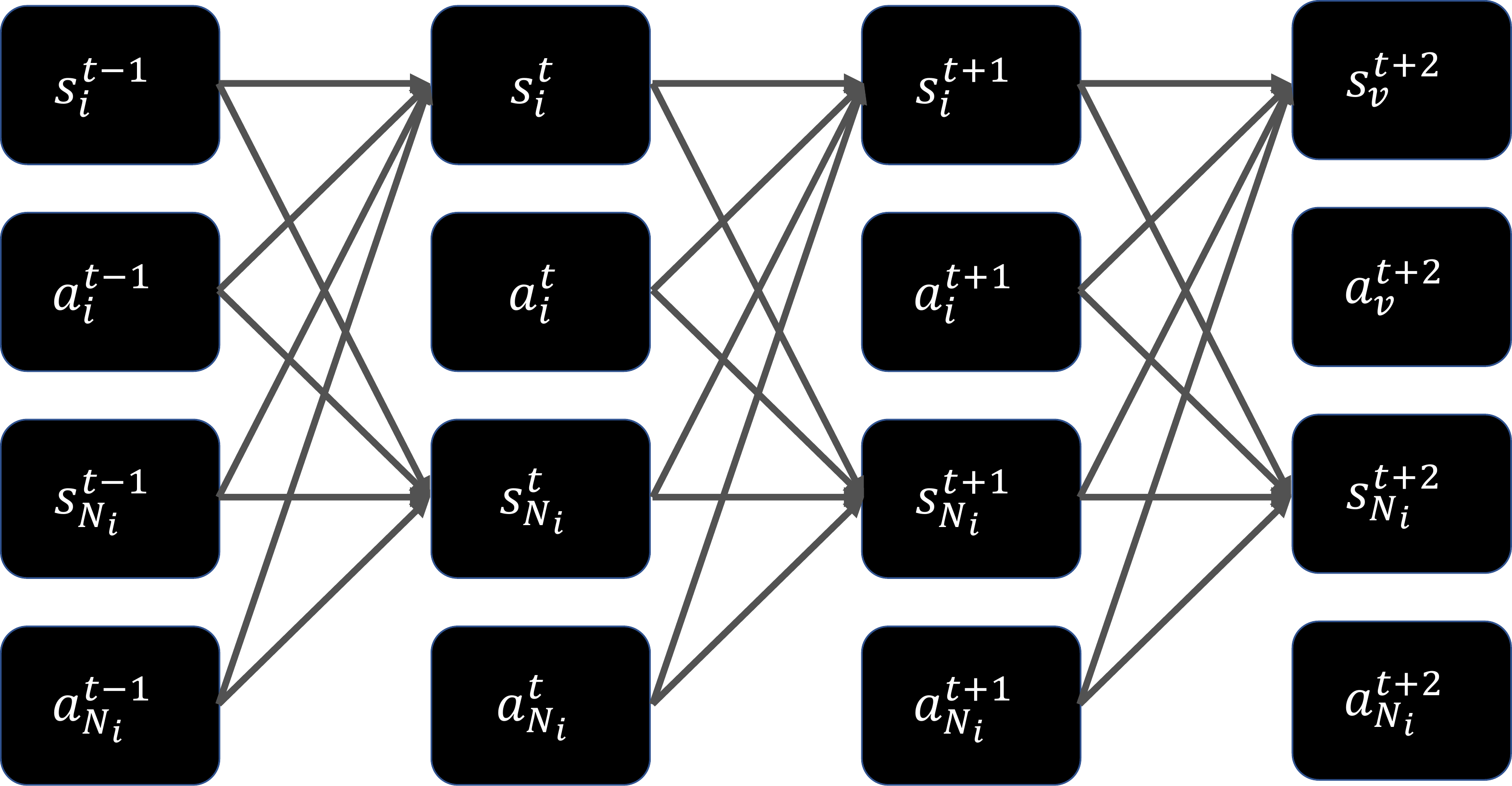}
\caption{Dependencies between states and actions at neighbouring intersections $i$ and $N_i$ across consecutive decision steps, with two-way traffic flow. A state represents a combination of local traffic conditions $\delta$ and initial phase conditions $\theta^0$ at a decision step. Two actions can be applied to the current phase at each decision step: \{\textit{extend, terminate}\}.}
\label{fig:gm}
\end{figure}

In this section, we provide a Decentralized Markov Decision Problem (Dec-MDP) representation for the traffic signal control problem over all intersections. It is represented as: 
\begin{align}
\left<{\cal I}, \times_i {\cal S}_i, \times_i {\cal A}_i, {\cal R} , \times_i {\cal P}_i, \right> \nonumber
\end{align}
\noindent {${\cal I}:$} Set of intersections. \\
\noindent ${\cal S}_i$: An intersection state, $s_i \in S_i$ represents a combination of  $\delta_i$ and $\theta^0_i$. Overall state, $s$ is jointly fully observable. \\
\noindent ${\cal A}_i$: The actions at each intersection at any decision epoch are {\em terminate} and {\em extend}. \\
\noindent ${\cal R}$: The reward function at an intersection is the sum of rewards (negative of delay/latency, ${\cal L}$) experienced by all vehicles: $ {\cal R} (s^t, a^t) = \sum_{i} {\cal R}_i (s^t_i, a^t_i) $ \\
\noindent ${\cal P}_i$: The transition function at an intersection is determined by $\{p_i^{\tau}\}_{\tau \in \Gamma}$ and phase model, $\phi_i$. 
Traffic observed at an intersection (part of the state space) is dependent only on traffic at previous time step at the same intersection and traffic moving to the intersection from neighbouring intersections, $N_i$. Figure~\ref{fig:gm} provides a graphical representation of these dependencies. The set of neighbouring intersections, $N_i$ is determined based on the planning horizon. This implies:
\begin{align}
 {\cal P}(s^{t+1} | s^t,a^t) =  \prod_{i \in {\cal I}} {\cal P}_i(s^{t+1}_i | s^t_i, s^t_{N_i}, a^t_i, a^t_{N_i}) \nonumber
 \end{align}
It should be noted that state of an intersection at a time step is not dependent on state or action of non-neighbour intersections at the same time step. This is crucial to value function of Dec-MDP becoming decomposable, as shown in the following proposition. 

\begin{prop}
For a Decentralized Markov Decision Problem (Dec-MDP) representing the distributed traffic signal control problem in synchronized settings, joint value function is decomposable over value of individual intersections, i.e., :
{\begin{align}
V^t(s^t) &= \sum_{i} V_i^t(s_i^t), \text{ where}\nonumber\\
V_i^t(s_i^t) &=  R_i(s_i^t, a_i^t) + \sum_{s_i^{t+1}} P(s_i^{t+1} | s_i^t, s_{N_i}^{t},a_i^t,a_{N_i}^{t}) V^{t+1}_i(s_i^{t+1}) \nonumber
\end{align}}
\label{prop:1}
\end{prop}
\noindent \textbf{Proof Sketch.} For the last time step, $H-1$, the proposition holds as the reward function is decomposable. Let us assume it holds for a time step $n = m+1$, then value function at $m$ is given by
{\begin{align}
V^{m}(s^{m})  &= \sum_{i} R_{i}(s_i^{m},a_i^{m}) + \sum_{s^{m+1}} P(s^{n} | a^m, s^{m}) V^{n}(s^{n}) \nonumber \\
&= \sum_{i} R_{i}(s_i^{m},a_i^{m})  + \sum_{s^m} \Big [ \prod_{i \in I} P(s_i^{n} | s_{i}^{m},s_{N_{i}}^{m},a_i^{m}, a_{N_i}^{m}) \Big ] V^{n}(s^{n}) \nonumber\\
\intertext{Since $V^{n}(s^{n})$ is decomposable by assumption, we have:}
&= \sum_{i} R_{i}(s_i^{m},a_i^{m})  +  \nonumber\\
&\hspace{0.3in} \sum_{s^{m+1}} \Big [ \prod_{i \in N} P(s_i^{m+1} | s_{i}^{m},s_{N_{i}}^{m},a_i^{m},a_{N_i}^{m}) \Big ] \sum_{i} V_i^{m+1}(s_i^{m+1}) \nonumber\\
&= \sum_{i} \sum_{i} R_{i}(s_i^{m},a_i^{m})  + \sum_{s^{n}} \sum_{i} P(s_i^{n} | s_{i}^{m},s_{N_{i}}^{m},a_i^{m},a_{N_i}^{m})V_i^{n}(s_i^{n}) \nonumber\\
&= \sum_{i} \Big[ R_{i}(s_i^{m},a_i^{m})  + \sum_{s_i^{n}} P(s_i^{n} | s_{i}^{m},s_{N_{i}}^{m},a_i^{m},a_{N_i}^{m})V_i^{n}(s_i^{n}) \Big] \nonumber\\
&= \sum_i V_i^m(s_i^m) \nonumber \hfill \quad \blacksquare
\end{align}}

Each individual intersection can therefore plan independently as long as it has information on state and action of current and neighbouring intersections. 
While Proposition~\ref{prop:1} provides a proof for distributing the traffic control problem the way it is described in  Section~\ref{sec:model},  it should however be noted that this proof assumes representation as a Dec-MDP where decision making at traffic intersections is synchronized. In reality, that may or may not always be feasible. Therefore, the above serves as a theoretical intuition and not a concrete proof for the real world scenario of asynchronous decision making at intersections. 

\section{Background: SURTRAC}

Scalable Urban Traffic Control (SURTRAC) is a recently developed real-time, distributed traffic signal control system which uses a schedule-driven approach to traffic signal control. \citep{xie2012schedule} formulate the traffic signal control problem as a single-machine scheduling problem, where each intersection is treated as a machine, and clusters of oncoming vehicles along competing routes are treated as jobs to be scheduled.
\begin{figure}
	\centering
	\subfloat[]{
		\includegraphics[height=1.7in,keepaspectratio]{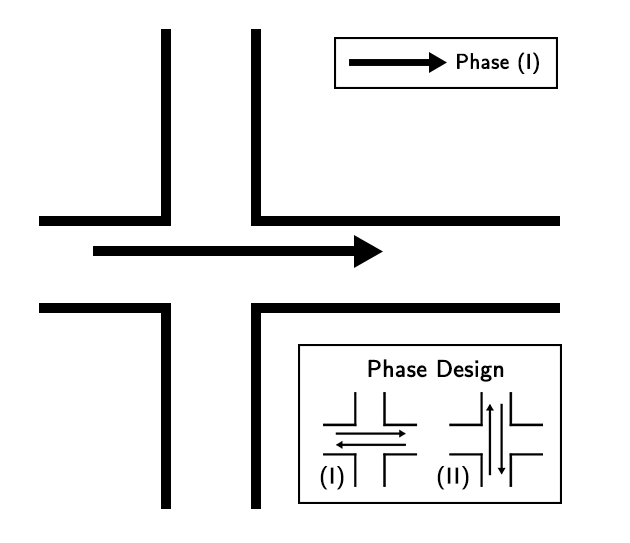}
		\label{fig:no-uncertainty}
	}
	\subfloat[]{
		\includegraphics[height=1.7in,keepaspectratio]{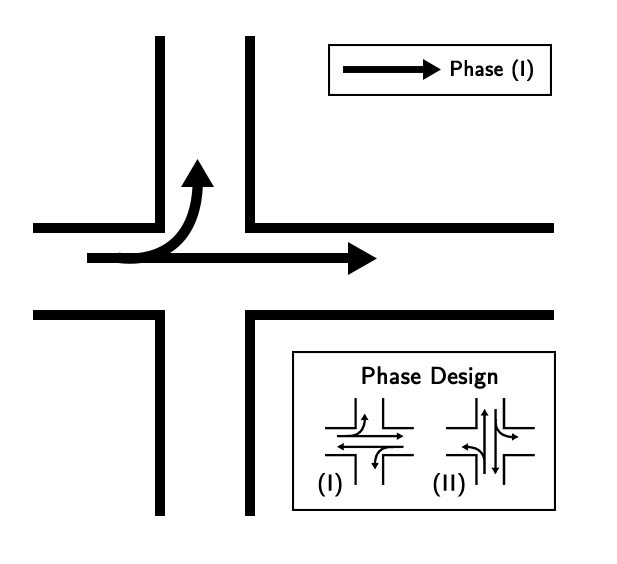}
		\label{fig:communication-uncertainty}
	}
	% Vehicles on an entry road can turn on to multiple exit edges. Projected traffic demand on these exit edges is uncertain, but is required for communication with neighbouring intersections, which use it to expand their observation horizons.
	\subfloat[]{
		\includegraphics[height=1.7in,keepaspectratio]{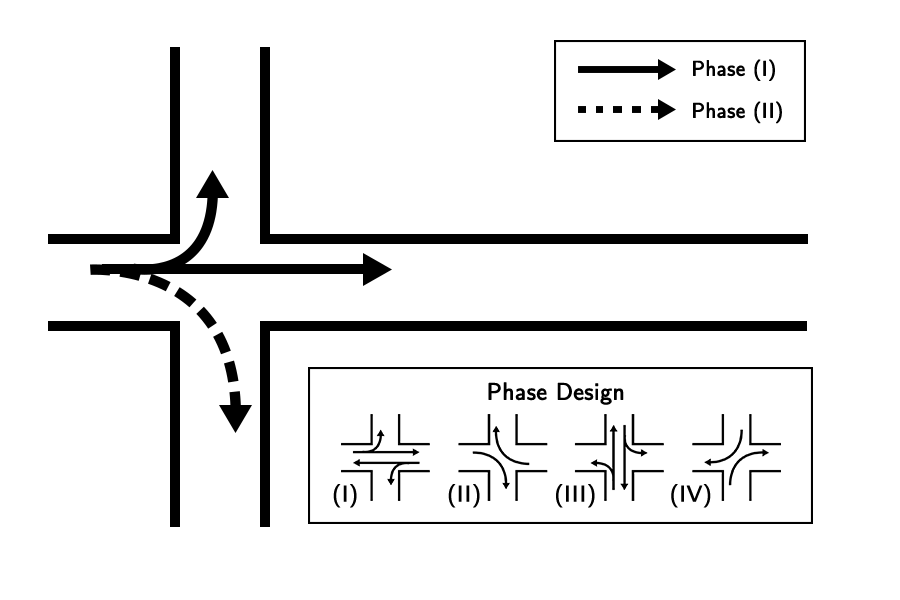}
		\label{fig:planning-uncertainty}
		% Vehicles on an entry road can exit the intersection (are serviced in) in competing phases. Traffic demand along these phases is uncertain, but is required for appropriate allocation of green time to the phases.
	}
	\caption{Manifestations of vehicle-turn-induced uncertainty. (a) No uncertainty; (b) Intra-phase uncertainty: Uncertainty in communication between neighbouring intersections (based on an estimate of the traffic on each exit edge) ; (c) Intra and inter-phase uncertainty: Uncertainty in planning (based on an estimate of the stochastic per-phase demand) and communication.}
\end{figure}
To ensure scalability, SURTRAC computes a schedule that minimizes delay for expected number of vehicles (instead of expected delay), i.e.:
{\begin{align}
\min_{\pi_i} {\cal L} \Big(\mathbb{E}_{p_i,p_{N_i}}[\delta_{i}^0], \ldots, \mathbb{E}_{p_i,p_{N_i}}[\delta_{i}^{|\phi_i|-1}] \hspace{0.05in} \boldsymbol{|} \hspace{0.05in} \delta_i,\delta_{N_i}, \lambda_i, \pi_i \Big)  \label{eqn:surtrac}
\end{align}}
Uncertainty associated with vehicle turn movements is the key source of uncertainty which results in the real objective, expected delay (of Equation~\ref{eqn:tuseract}) to be different from SURTRAC's objective, delay for expected number of vehicles (of Equation~\ref{eqn:surtrac}). This uncertainty manifests itself as:
\squishlist
	\item Intra-phase uncertainty: This arises when vehicles on an entry edge can exit the intersection in the same phase, but on to different exit edges (as shown in Figure~\ref{fig:communication-uncertainty} ). This has an impact on traffic outflows $\delta_{N_i}$ communicated to intersection $i$.
	\item Intra and Inter-phase uncertainty: This arises when vehicles on an entry edge can exit the intersection in multiple, competing phases (as shown in Figure~\ref{fig:planning-uncertainty}). This has a direct not only on $\delta_{N_i}$, but also on per-phase traffic flow estimation at intersection $i$, i.e., $\delta_i$.
\squishend

\section{Turn-Sample-Based Real-Time Traffic Signal Control (TuSeRACT)}

We now propose a real-time, distributed, schedule-driven, sample-based approach to traffic signal control. Our approach, named TuSeRACT is based on the framework defined by SURTRAC in that it is a distributed approach that treats traffic signal control as a cluster scheduling problem. However, it aims to minimize expected delay (and not minimize delay for expected traffic as in SURTRAC) for outgoing traffic over all intersections.

In order to handle the intra and inter-phase uncertainty, we employ \textit{Sample Average Approximation} (SAA)~\citep{kleywegt2002sample}.  SAA allows us to approximate the expectation of delay over intra and inter-phase uncertainty by averaging multiple realizations of traffic flow along different phases. Formally, we approximate the expectation of delay $\mathcal{L}$ caused by a signal timing plan $\pi_i$ by averaging the delay it causes over a sample set, $\Xi_i$ of traffic realizations in different phases. Each realization/sample, $\xi \in \Xi_i$ is represented as:
$${\xi = \{\delta_i^0(\xi), \cdots, \delta_i^{|\phi_i|-1}(\xi)\}}$$
This transforms the original stochastic optimization problem of Equation~\ref{eqn:tuseract} into the following deterministic problem:
{\begin{equation}
\min_{\pi_i} \frac{1}{|\Xi_i|} \sum_{\xi \in \Xi_i} \mathcal{L} (\xi \hspace{0.05in}| \hspace{0.05in} \delta_i, \delta_{N_i}, \lambda_i, \pi_i)\label{eq:saa}
\end{equation}}

We solve this resulting sampling-based deterministic optimization problem using constraint programming. At each decision step, each intersection independently samples turn movements for observed vehicles, and computes a signal timing plan which minimizes the average delay across these samples. We now describe the various components of our sampling-based constraint programming approach.

\subsection{Traffic Representation and Sampling}

Our approach represents oncoming traffic as clusters of vehicles, and aggregates clusters into samples of per-phase cluster sequences (inflows), which are then used to compute a delay-minimizing schedule for the observed clusters. We use sensed vehicular data from each entry road of the intersection. For each observed vehicle on entry edge $e$, we independently sample an exit edge $f$ according to the given turn probabilities, $p_{i}^{\tau}$ (where $\tau = \left<e,f\right>$). A complete sample drawn by each intersection at each decision point is the vector of sampled exit roads for all observed vehicles at an intersection. Since the intersection phase design $\phi_i$ maps each permissible turn to a single phase, sampling an exit road (or a turn) for a vehicle is equivalent to sampling the phase in which the vehicle will leave the intersection. Post sampling, vehicles are aggregated into per-phase cluster sequences by proximity, based on sampled exit phases and estimated arrival times at the stop line of the intersection. During aggregation, we record cluster composition $\gamma$ which stores estimated arrival time $a$ and sampled exit edge $f$ for each vehicle in the cluster. This sampling and clustering process is repeated $|\Xi_i|$ times at each decision point.

\subsection{Constrained Optimization}

As described, our approach to the cluster scheduling problem under uncertainty is based on sample average approximation, where we aim to compute a signal timing plan, $\pi_i$ which minimizes the average delay across sampled traffic inflows. We formulate this deterministic, sample-based cluster scheduling problem (formalized in Equation~\ref{eq:saa}) as a constraint program, and solve it using the IBM ILOG CP Optimizer. 

\noindent The key \textbf{inputs} to constraint program are as follows: 
\squishlist
\item[$\boldsymbol{\Xi_i}$:] Sampled inflows $\xi \in \Xi_i$ include clusters on each $\psi^k$, i.e.,
{\begin{align}
 \xi &= \{\delta_i^0(\xi), \cdots, \delta_i^{|\phi_i|-1}(\xi)\} \nonumber\\ 
\delta^k_i(\xi) &= \Big< c^{k}_{i,1} (\xi), \dots, c^{k}_{i,q} (\xi)\Big>
\end{align}}
  $c^k_{i,q}(\xi)$ represents the cluster at $q$ position on phase $k$ at intersection $i$ according to sample $\xi$. It is characterized by:
$$\{|c^k_{i,q}(\xi)|, a^k_{i,q}(\xi), l^k_{i,q}(\xi), \gamma^k_{i,q}(\xi)\}$$
 $|c^k_{i,q}(\xi)|$ is the number of vehicles in the cluster, $a^k_{i,q}(\xi)$ is the arrival time at intersection $i$ (with $a^k_{i,q}(\xi) \geq a^{k-1}_{i,q}(\xi)$), $l^{k}_{i,q}(\xi)$ is the length of the cluster and $\gamma^k_{i,q}(\xi)$ is the cluster composition.
\item[$\boldsymbol{H}:$] Planning horizon or equivalently the number of cycles. A cycle represents one complete sequence of all phases. 
\item[$\boldsymbol{\left<\lambda_i, \delta_{N_i}\right>}$:] Distributed real-time traffic signal control inputs.\\
\squishend

\noindent \textbf{Decision Variables:} We model our decision variables using the notion of \textit{interval variables}, a CP Optimizer feature which allows us to intuitively model scheduling problems in terms of intervals of time. An interval variable $\iota$ is an interval of time $[start(\iota), end(\iota))$, where $start(\iota)$ and $end(\iota)$ are integral. $length(\iota)$ is defined as $end(\iota) - start(\iota)$. An interval variable can be optional, i.e. it may or may not be present in the solution.

The key decision variables are as follows:
(i) {\em Traffic signal timing plan} or phase intervals, $\pi^{k,r}_i$, which represents interval for phase $k$ in cycle $r$ for intersection $i$.
(ii) {\em Outgoing cluster intervals,}$O^{k,r}_{i,q}(\xi)$: Since we assume that clusters are divisible, fragments of a cluster may be scheduled in any cycle.  $O^{k,r}_{i,q}(\xi)$  represents the fragment of cluster $c_{i,q}^k(\xi)$ scheduled in cycle $r$ (if any).  $start(O^{k,r}_{i,q}(\xi))$ represents the time at which the cluster fragment starts leaving the intersection by crossing the stop line.  The absence of $O^{k,r}_{i,q}(\xi)$ implies that no fragment of $c_{i,q}^k$ is scheduled in cycle $r$.

\noindent\textbf{Constraints:}
We now describe intuitively and formally the main constraints that are required to compute an optimal traffic signal control plan.  The computed signal timing plan, $\pi_i$ and outgoing cluster departures are constrained by the the initial conditions at the decision point, the observed traffic inflow and the distributed traffic signal control inputs. Here are the main constraints: \\
\noindent \textit{Phase duration}: The duration of each phase is constrained by a minimum green and a maximum green time predefined by the signal timing specification.
	{\begin{equation}
	length(\pi_{i}^{k,r}) \in [G_{min}^k, G_{max}^k] \quad \forall r, \psi^k \in \phi_i
	\end{equation}}
\noindent \textit{Phase order and inter-green time}: In each cycle, the phases $\phi_i$ must be given right-of-way in a fixed order per the signal timing specification, and a fixed intergreen time must be applied between consecutive phases.
	{\begin{equation}
	start(\pi_{i}^{k,r}) = end(\pi^{k-1,r}_i) + Y^{k-1} \quad \forall \psi^k \in \phi_i, k \neq 1, r 
	\end{equation}}
\noindent \textit{Cycle order}: The cycles occur in a fixed order, and a fixed inter-green time is applied between consecutive cycles.
	{\begin{equation}
	start(\pi^{1,r}_i) = end(\pi^{|\phi_i-1|,r-1}_i) + Y^{|\phi_i-1|} \quad \forall r \neq 1
	\end{equation}}
\noindent \textit{Initial conditions}: The current phase $\psi^0$ at decision time $t$ determines the feasibility of extension of the current signal timing plan. We formulate this by constraining the start and end times of $\psi^0$ in the current cycle $r = 1$.
	{\begin{align}
	start(\pi_i^{0,1})  &= t - g \hspace{0.1in};\hspace{0.1in} end(\pi_i^{0,1}) &\ge t
	\end{align}}
\noindent \textit{Handling cluster fragments}: We assume that clusters are divisible, and that fragments of a single cluster can be scheduled in different cycles. The length of each fragment of a cluster is constrained by the length of the cluster:
	{\begin{equation}
	length(O^{k,r}_{i,q}(\xi)) \in [1, l^k_{i,q}(\xi)] \quad \forall \xi \in \Xi_i, \psi^k \in \phi_i, r, q
	\end{equation}	}
	We ensure that each cluster completely leaves the intersection in all cycles:
	{\begin{equation}
	\sum_{r} length(O^{k,r}_{i,q}(\xi)) = l^k_{i,q}(\xi)\quad \forall \xi \in \Xi_i, \psi^k \in \phi_i, r, q
	\end{equation}}
\noindent \textit{Cluster departure}: % Better name?
	Outgoing clusters can be scheduled to exit the intersection only after they arrive at the stop line:
	{\begin{equation}
	start(O^{k,r}_{i,q}(\xi)) \ge  l^k_{i,q}(\xi) \quad \forall \xi \in \Xi_i, k,r, q
	\end{equation}}
	Fragment $O^{k,r}_{i,q}(\xi)$, if scheduled, must be scheduled in the appropriate phase $\pi_{i}^{k,r}$:
	{\begin{align}
	start(O^{k,r}_{i,q}(\xi)) &\ge start(\pi_{i}^{k,r}) \quad \forall \xi \in \Xi_i, \psi^k \in \phi_i,r, q \\
	end(O^{k,r}_{i,q}(\xi)) &\le end(\pi_{i}^{k,r}) \quad \forall \xi \in \Xi_i, \psi^k \in \phi_i,r, q
	\end{align}}
\noindent \textit{Cluster precedence among clusters in the same phase}: Any fragment of cluster $c^{k}_{i,q}$ can be scheduled only all fragments of $c_{i,q-1}^{k}$ have been scheduled. In other words, $c^{k}_{i,q}$ can be scheduled only after $c_{i,q-1}^{k}$ has completely exited the intersection.
	{\begin{align}
	&presenceOf(O^{k,r}_{i,q}(\xi)) \rightarrow not(presenceOf(O^{k,r'}_{i,q-1}(\xi)) \nonumber\\
	&\quad \forall \xi \in \Xi_i, k \neq 0, r, r' \in \{r+1, r+2, \dots \}, q
	\end{align}
	\noindent($r'$ is any succeeding cycle)
	\begin{equation}
	end(O^{k,r}_{i,q-1}(\xi)) \leq start(O_{i,q}^{k,r})
	\end{equation}}
	$\forall \xi \in \Xi_i, \psi^k \in \phi_i, q \neq 0, r$\\
\noindent \textbf{Objective:} We compute a single signal timing plan, $\pi_i$ which minimizes the cumulative waiting time across the vehicles in all the inflow samples:
{\begin{equation}
\min \displaystyle\sum_{\xi,k,q,r} \Big(start(O^{k,r}_{i,q}(\xi)) - a^k_{i,q}(\xi)\Big) \cdot |c^k_{i,q}(\xi)| \cdot \frac{size(O^{k,r}_{i,q}(\xi))}{l^k_{i,q}(\xi)}
\end{equation}}

\noindent $\Big(start(O^{k,r}_{i,q}(\xi)) - a^k_{i,q}(\xi)\Big)$ represents the delay incurred by cluster fragment $O^{k,r}_{i,q}(\xi)$, and $ |c^k_{i,q}(\xi)| \cdot \frac{size(O^{k,r}_{i,q}(\xi))}{l^k_{i,q}(\xi)}$ represents the number of vehicles in that fragment.

\subsection{Sample-Based Communication}
Like SURTRAC, our approach to traffic signal control is that of distributed, local planning with communication with neighbours. Intersections independently compute signal timing plans for traffic in their respective observation horizons and communicate projected outflows to neighbouring intersections to expand their observation horizon. Limiting communication to immediate neighbours makes both approaches scalable to large road networks.

Our approach is to communicate samples of vehicle departure times to neighbouring intersections along the sampled exit roads as opposed to communicating a single expected outflow cluster sequence along each exit road. At each intersection $i$, vehicle departure times for each of the $|\Xi_i|$ samples are computed using $\pi_{i}$ and communicated to the appropriate neighbours $N_i$ according to the sampled exit edges. At $N_i$, the set of received non-local vehicle arrival times is appended to the locally observed temporal arrival distribution in order the increase the observation horizon. Exit edges are then sampled for all vehicles in this extended observation horizon.

\subsection{Guided Search}

In order to exploit situations where traffic patterns do not change drastically between consecutive decision points, we also provide a heuristic to find better solutions in the limited compute time available. The key idea here is to use the signal timing plan generated at the previous decision point as an initial starting point for the CP Optimizer search at the current decision point. If traffic patterns do not change drastically, the previously computed solution can serve as a very good solution for the current decision step. We refer to this heuristic as \textit{guided search} (gs).

\section{Performance Evaluation}

In this section, we provide a comparison against the leading approach for real-time distributed traffic signal control, SURTRAC. We use the same networks from SURTRAC research~\citep{xie2012schedule,xie2012coordination} to ensure no advantage to our work. Specifically, we experiment with the following synthetic road networks: (1) an isolated intersection (2) a 1x5 grid network (3) a 5x5 grid network.  All simulations are run on Simulation of Urban MObility \citep{behrisch2011sumo}, an open source traffic simulation package. 

We implement and evaluate two versions of TuSeRACT: Sample-based cluster scheduling without communication with neighbouring intersections (UTuS) and sample-based scheduling with communication with neighbouring intersections (CTuS). We compare against two versions of SURTRAC baselines: SURTRAC without communication among neighbouring intersections (USUR) \citep{xie2012schedule} and SURTRAC with communication between neighbours (CSUR) \citep{xie2012coordination}. We use the IBM ILOG CP Optimizer (single thread, 5s solver compute time limit unless otherwise stated) to solve the sample-based scheduling problem at every decision point. Mean vehicle waiting time is used as an indicator of solution quality.

We vary road lengths (observation horizon) and phase designs across networks in order to evaluate our approach in a variety of scenarios. For each network, we define a traffic demand profile which specifies how the total traffic demand is distributed over the input edges in the network over time. For each network, we run simulations on three traffic demand levels corresponding to low, medium and high traffic level for that network.

Given pre-defined turn proportions which remain static throughout the simulation and an incoming traffic demand profile, we generate a fixed set of vehicle routes for multiple demand levels for each network. Traffic is generated for 15 minutes and the problem horizon extended till all vehicles have cleared the network as in \citep{guilliard2016non}. Each of these set of routes is used to generate 20 test instances (where routes are fixed, but vehicle arrival times vary), which we then evaluate our approach on. We evaluate our approach on each instance with 5 independent and identically distributed sample sets, which we generate offline using the fixed routes and the preset turn proportions. For each sample set, we run TuSeRACT with \{1, 5, 10, 20, 30\} samples. This amounts to 100 runs of TuSeRACT for each demand level and sample count.

We generate samples offline to study the effect of adding additional samples to an existing sample set. In practice, vehicle turns would be sampled online at each decision step.

As in \citep{xie2012schedule}, we assume that vehicles travel at the constant speed of 10 m/s and queued vehicles are discharged at a saturation flow rate of $N_{lane}/2.5$ vehicles / second after a startup lost time of 3.5s, where $N_{lane}$ is the number of lanes along the incoming road. We assume all oncoming vehicles are passenger vehicles with length = 5m. We cluster them at a sampling interval $samp$ = 1s, and use a threshold of 3s to further aggregate clusters by proximity. For each phase across the simulations, we set $G^{min}$ to 5 seconds, $G^{max}$ to 55 seconds and $Y$ to 5 seconds. We use an optimization horizon of 3 cycles (including the current cycle).

The time resolution used by our implementation of SURTRAC is 0.5s while that used by our approach is 1s. We use an optimization horizon of 3 cycles (including the current cycle) to limit schedules computed by TuSeRACT.

For both approaches, we assume that:
\squishlist	
\item Vehicles can be detected exactly along the full observation horizon (usually the full incoming road segment in our cases). In other words, we ignore sensor detection error.
	\item Turn proportions remain static throughout a simulation and are known exactly by each intersection. In practice, these turn proportions are estimated as moving averages based on recently observed turn movements.
	\item Planning and communication are instantaneous. For both approaches, planning and communication time are counted outside the simulation and have no impact on the simulation. We make this simplifying assumption to focus solely on the solution quality of the two approaches. We discuss the trade-off between solution quality and real-time tractability in the context of both the approaches, but do not aim to address it in this paper. Here, we aim to investigate whether sampling-based traffic signal control can reduce waiting times when planning under turn-induced uncertainty.
\squishend

\begin{figure}[htbp]
	\centering
	\subfloat[] {
		\includegraphics[height=2in,keepaspectratio]{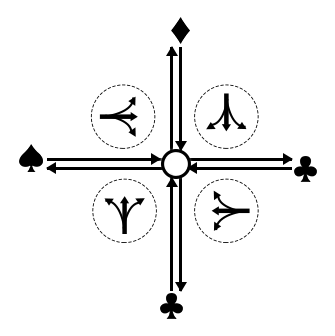}
		\label{fig:isolated-net}
	}
	
	\subfloat[]{
		\includegraphics[height=2in,keepaspectratio]{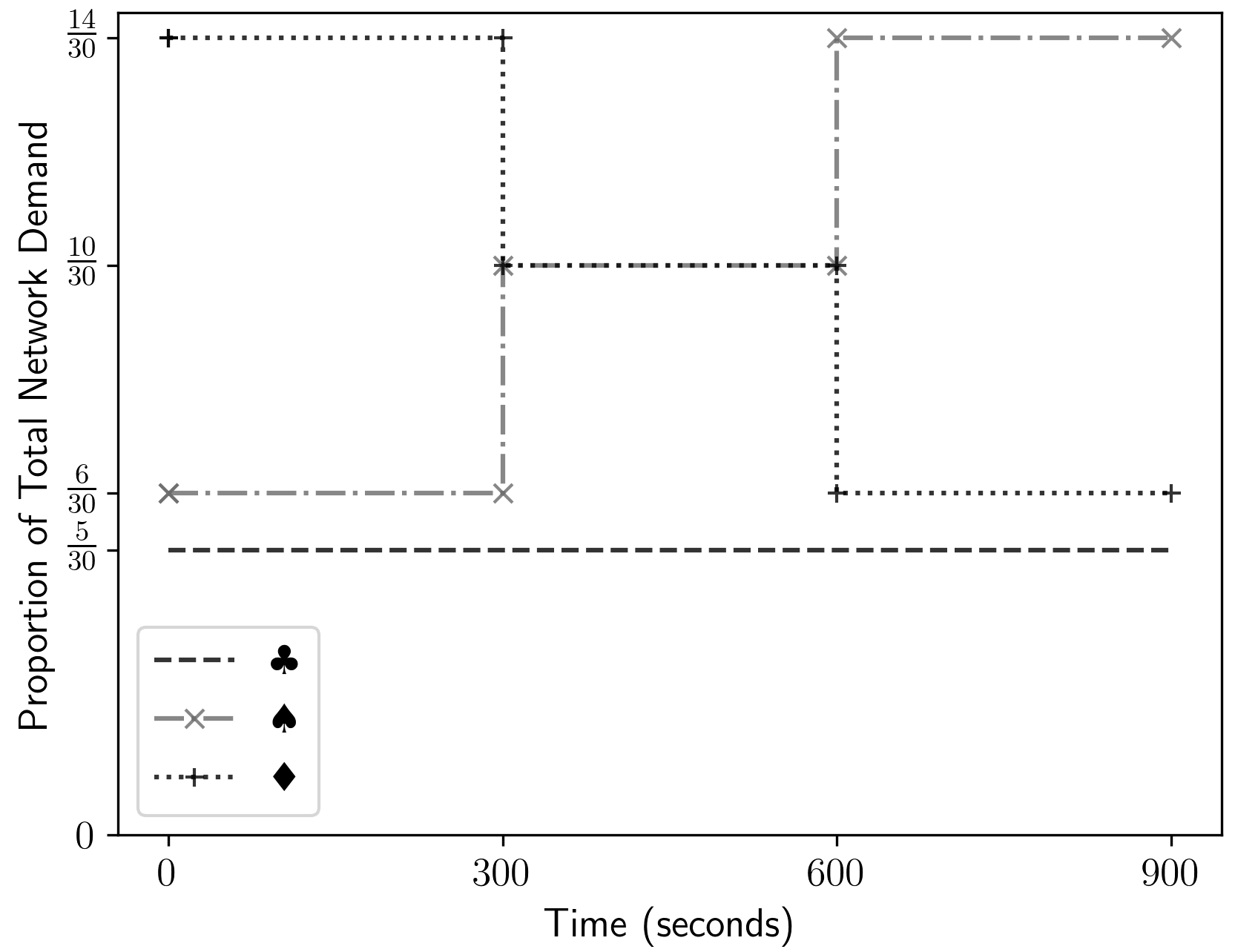}
		\label{fig:isolated-flow}
	}
	\caption{Experimental Setup for an isolated intersection. (a) Arrows indicate direction of vehicle movement and permissible turns are shown alongside  incoming approaches; (b) Demand flow profile}
\end{figure}

\begin{figure*}[htbp]
	\centering
	\includegraphics[height=2.8in,keepaspectratio]{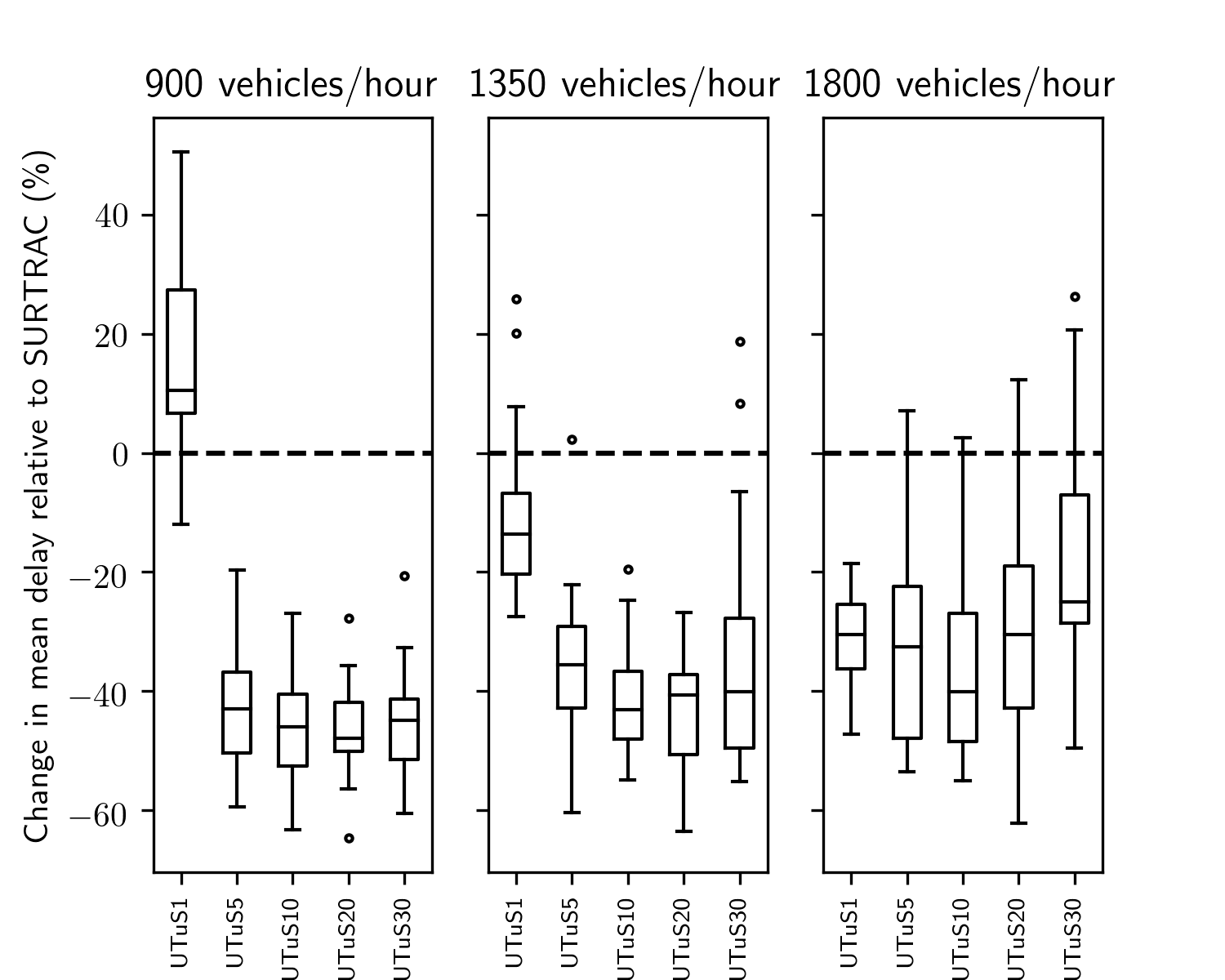}
	\label{fig:1x1-vs-usurtrac}
	\caption{Isolated intersection:  Performance gain achieved by UTuS over USUR for various sample counts and demand levels. Each boxplot summarizes 100 data points.}
\end{figure*}

\begin{table}[htbp]
	\centering
	\begin{tabular}{|c|c|c|}  
		\toprule
		Demand & \multicolumn{2}{c}{Change in mean delay over SURTRAC (\%)} \\
		(vph) & TuS10 & TuS10 + gs \\
		\midrule
		900 & -45.70 $\pm$ 8.99 & -54.39 $\pm$ 6.36 \\
		1350 & -43.45 $\pm$ 13.29 & -53.42 $\pm$ 8.66 \\
		1800 & -38.37 $\pm$ 17.94 & -50.29 $\pm$ 6.58 \\
		\bottomrule
	\end{tabular}
	\caption{Effect of guiding CP Optimizer search with the previous decision step's solution for an isolated intersection. For brevity, we report results for TuS with 10 samples.}
	\label{tab:isolated-reuse}
\end{table}

\subsection{Isolated Intersection}
We evaluate our approach on a four-phase (Figure~\ref{fig:phases}), two-lane, two-way single intersection (Figure~\ref{fig:isolated-net}). All incoming roads are 300m long (equivalent to a 30s observation horizon), and that vehicles can turn left, right or travel straight through from each incoming road. 60\% of the vehicles do not turn, and 20\% turn right and left respectively. We generate routes according to the demand profile in Figure~\ref{fig:isolated-flow}, where we show the distribution of the total demand over the incoming approaches over the traffic generation period. We use this profile to generate test cases for three demand levels (vehicles/hour or vph): \{900, 1350 and 1800\}.

To evaluate performance, we report the percentage change in mean waiting time relative to USUR, averaged over all test cases and sample sets (Figure~\ref{fig:1x1-vs-usurtrac}). Our experiments show that: (i) UTuS results in significantly lower (40\% on average) mean vehicle waiting times with respect to USUR across all demand levels. (ii) The number of samples required to produce this substantial improvement is small. Although a single sample is not sufficient across all demand levels, using 5 or more samples consistently provides a 35-45\% reduction in delay across all demand levels. (iii) Increasing the sample count beyond 10 does not significantly improve performance. In fact, this results in a performance drop for higher demand levels due to the tougher computational challenge (larger constrained optimization to be solved in 5 seconds) posed to UTuS (from both high demand and high sample counts).

To verify that this degradation is due to insufficient compute time, we also ran simulations on with a solver time limit of 30 seconds. The increased compute time indeed results in higher performance gains ($\sim$45\%) even for high demand (1800 vph) and high sample counts (20, 30). With a 5 second time limit, solution quality can be improved with guided search (Table~\ref{tab:isolated-reuse}).

\begin{figure}[htbp]
	\centering
	\subfloat[] {
		\includegraphics[height=1.5in,keepaspectratio]{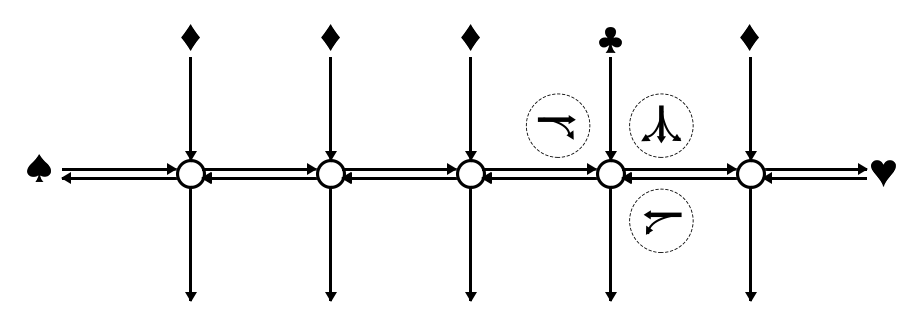}
		\label{fig:1x5}}
	
	\subfloat[] {
		\includegraphics[height=2in,keepaspectratio]{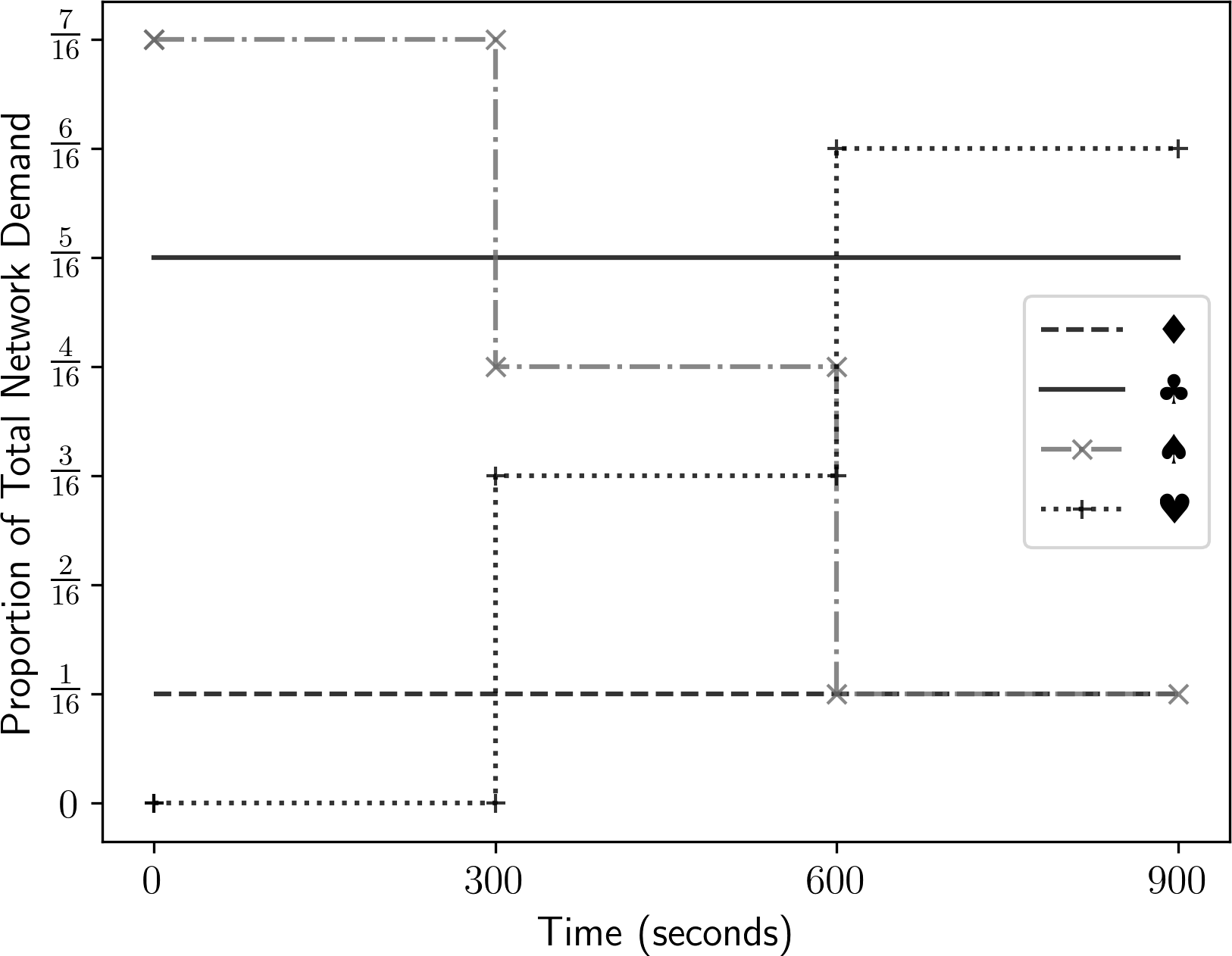}
		\label{fig:1x5-flow}}
	\caption{Arterial Network. (a) Permissible turn movements shown alongside respective incoming roads; (b) Demand flow profile.}
\end{figure}

\begin{figure*}[htbp]
	\centering
	\includegraphics[height=2.8in,keepaspectratio]{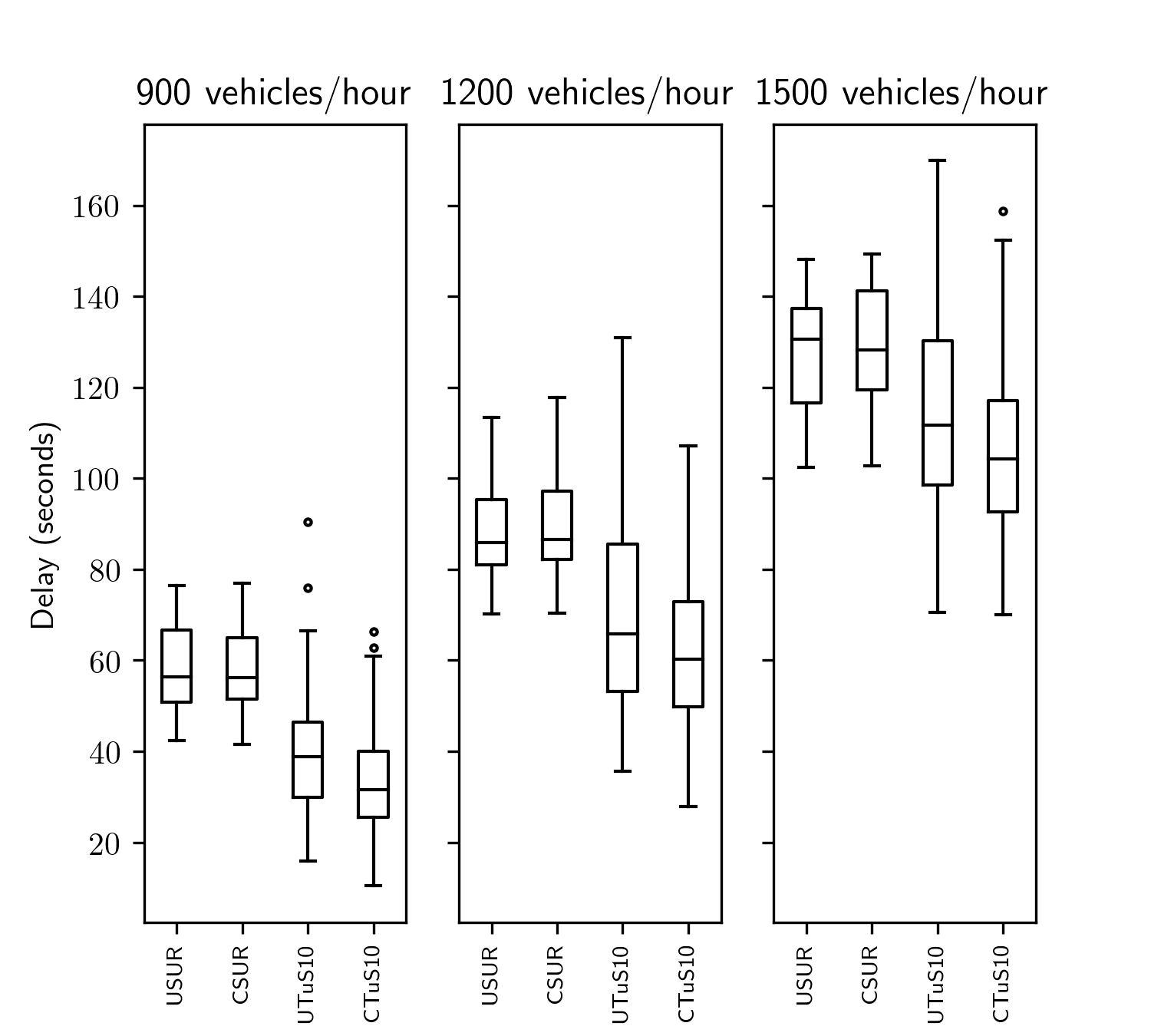}
	\caption{Arterial network: Absolute delays for coordinated and uncoordinated SUR and TuS10 for various demand levels. Each TuS boxplot summarizes 100 data points (over 20 test cases).}
	\label{fig:1x5-results}
\end{figure*}

{\begin{table}
		\centering
		\begin{tabular}{|c|c|c|c|}  
			\toprule
			Approach & \multicolumn{3}{c}{Change in mean delay over SURTRAC (\%)}  \\
			& 900 vph & 1200 vph & 1500 vph\\
			\midrule
			UTuS10 & -42.4 $\pm$ 10.3 &-24.0 $\pm$ 26.5 &-9.5 $\pm$ 11.1 \\
			UTuS10 + gs & -38.8 $\pm$ 16.2 & -26.6 $\pm$ 12.0 & -15.8 $\pm$ 7.1 \\
			CTuS10 & -41.9 $\pm$ 17.7 & -33.7 $\pm$ 15.8 & -15.3 $\pm$ 12.8 \\
			CTuS10 + gs& -36.0 $\pm$ 19.3 & -21.7 $\pm$ 17.8 & -5.8 $\pm$ 13.3\\
			\bottomrule
		\end{tabular}
		\caption{Effect of guiding CP Optimizer search with the previous decision step's solution for an arterial network.}
		\label{tab:1x5-reuse}
\end{table}}

\subsection{Arterial Network}

Next, we evaluate our approach on a 5-intersection arterial network (Figure~\ref{fig:1x5}). All road lengths are 250m, equivalent to a 25s observation horizon. As shown, we permit turning only on one three-phase bottleneck intersection. All other intersections are all two-phase intersections. For approaches that allow one turn, traffic turns according to (80\% through, 20\% left or right). For approaches that allow both left and right turns, traffic turns as (65\% through, 15\% left, 20\% right). For communication, we set the horizon extension to 20s. Traffic is generated for the demand levels (vph) \{900, 1200, 1500\} according to the demand profile shown in Figure~\ref{fig:1x5-flow}. This scenario tests our approach in networks with low uncertainty.

To allow comparison between SURTRAC and TuSeRACT, but also coordinated (CTuS) and uncoordinated (UTuS) approaches, we plot the absolute delays across approaches (Figure~\ref{fig:1x5-results}). For brevity, we only report results for TuSeRACT with 10 samples. Our results show that TuSeRACT is able to perform at par with SURTRAC, and even reduce mean waiting times (by 10-40\%) in scenarios with low uncertainty. However, for both approaches, coordinated approaches do not significantly outperform uncoordinated approaches. This could be due to the fact that: (i) The observation horizon is sufficiently long and non-local information does not provide significant advantage during planning. (ii) The bottleneck intersection causes queue spillover which propagates to neighbouring intersections. \citep{xie2012coordination} proposes a spillover prevention strategy, but we do not explore it here. 

We also note that guided search does not consistently improve solution quality here (Table~\ref{tab:1x5-reuse}). This could be due to accelerated movement of traffic through non-bottleneck intersections, which results in quickly changing traffic inflow patterns which render re-use of previous solutions disadvantageous. These require further investigation.

\begin{figure}[htbp]
	
	\centering
	\subfloat[]{
		\includegraphics[height=2.2in,keepaspectratio]{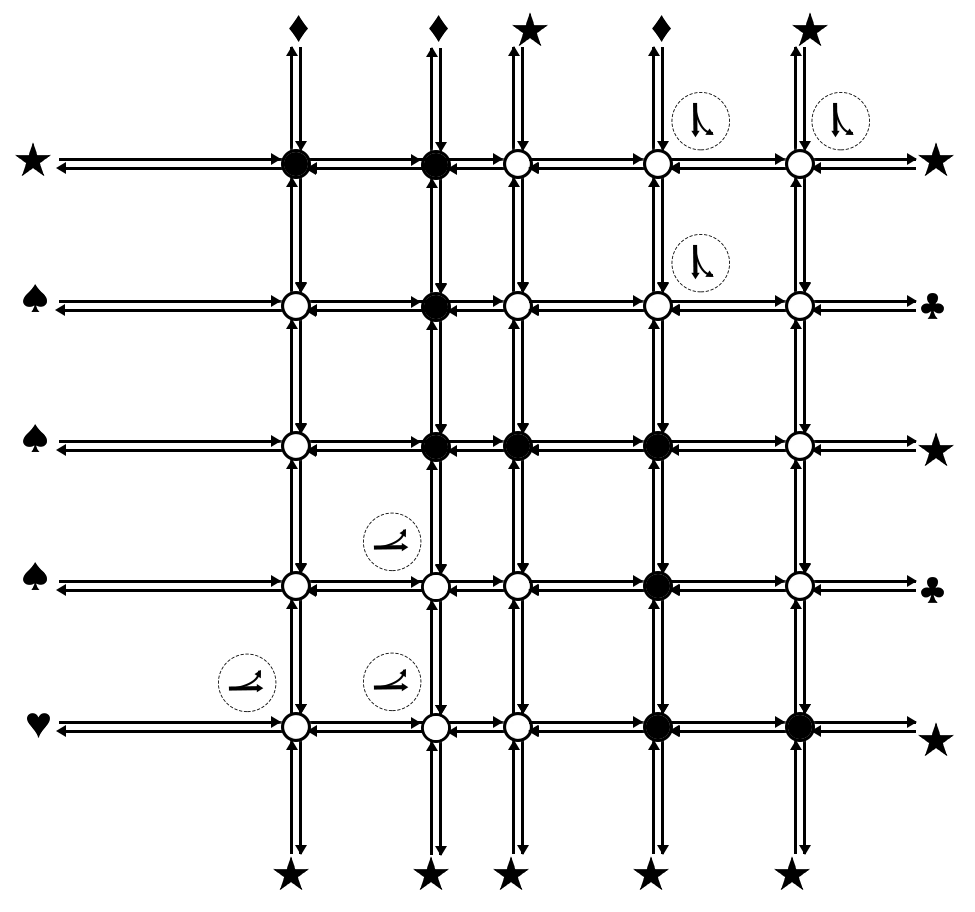}
		\label{fig:5x5}}
	
	\subfloat[] {
		\includegraphics[height=2in,keepaspectratio]{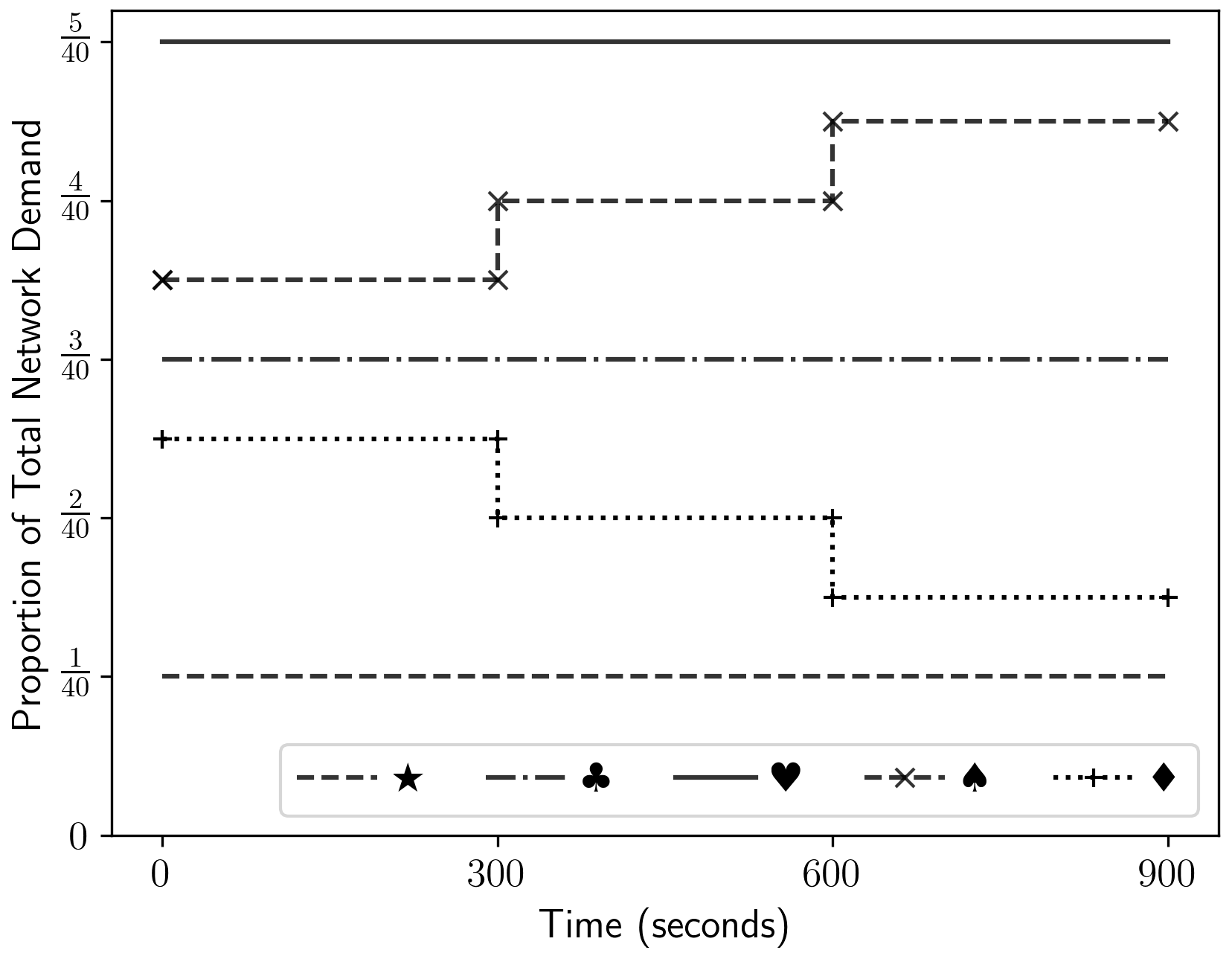}
		\label{fig:5x5-flow}}
	\caption{Experimental Setup for a 5x5 Grid Network. (a) Darkened circles represent 4-phase intersections where through, left and right turns are permissible from each incoming road. The other intersections are 2-phase intersections. Additional turn movements (if any) are shown alongside the respective roads; (b) Demand flow profile.}
\end{figure}

\begin{figure*}[htbp]
	\centering
	\includegraphics[height=2.8in,keepaspectratio]{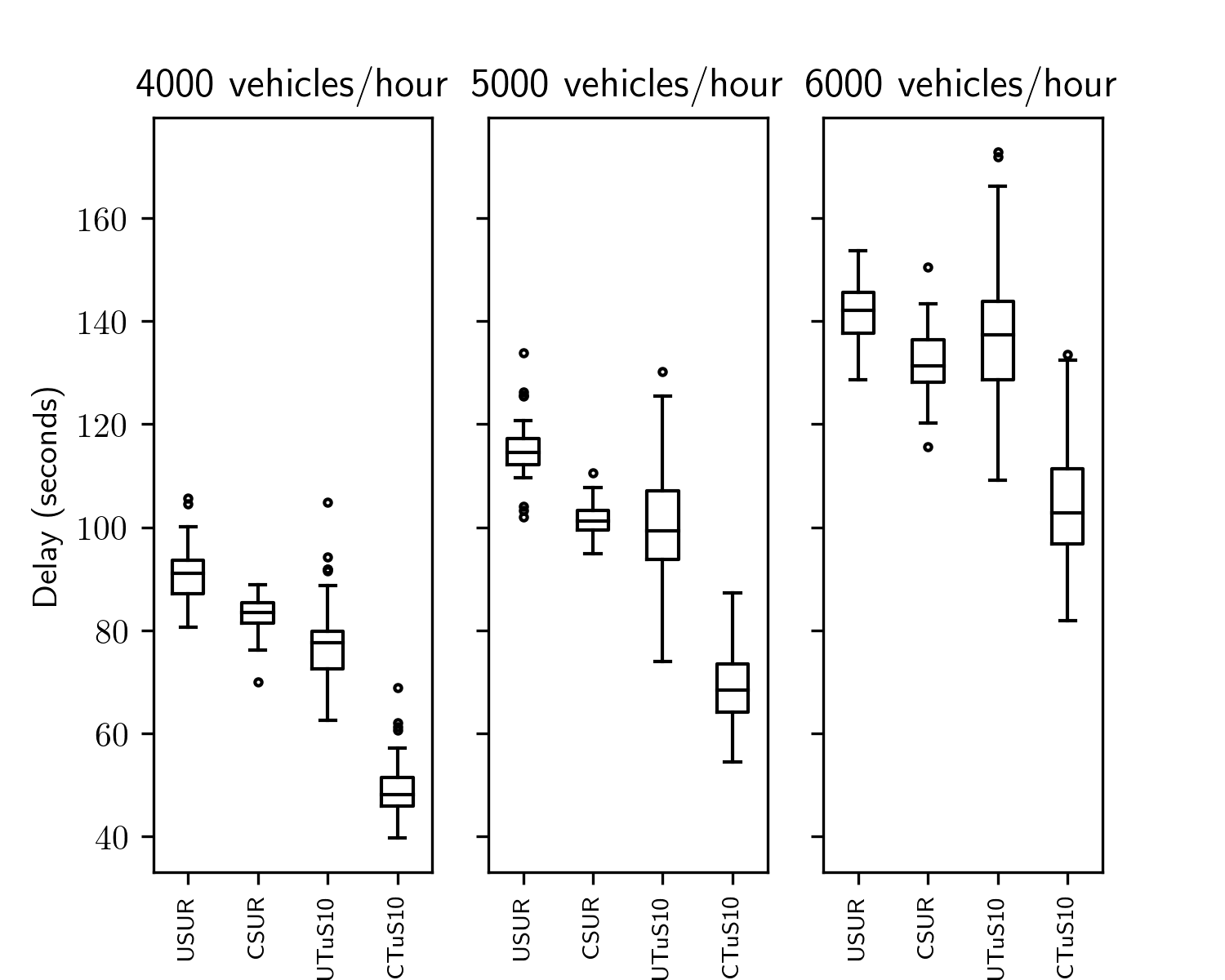}
	\caption{5x5 grid network: Absolute delays for coordinated and uncoordinated SUR and TuS10 for various demand levels. Each TuS boxplot summarizes 100 data points (over 20 test cases).}
	\label{fig:5x5-results}
\end{figure*}

\subsection{5x5 Grid Network}
Finally, we evaluate our performance on a synthetic 5x5 network shown in Figure~\ref{fig:5x5}, which also indicates the road lengths in the network. All road lengths are 75m (7.5s observation horizon), except for one set of inner roads (25m) and one set of boundary edges (150m). For approaches which allow one turn, traffic turns according to (80\% through, 20\% left or right) and for those which allow both left and right turns, it turns according to (65\% through, 15\% left, 20\% right). We generate demand for levels (vph) \{4000, 5000, 6000\} for this network according to the demand profile shown in Figure~\ref{fig:5x5-flow}. Since the intersections are closely placed (as in a typical urban network), this represents a situation which would benefit from coordinated planning.

Figure~\ref{fig:5x5-results} shows the results of our experiments. At 10 samples, UTuS provides a 7-17\% reduction in delay over USUR and CTuS provides a 19-40\% delay over CSUR. We also observe that the performance gain can be significantly improved (up to a consistent 40-50\%) with guided search (Table~\ref{tab:5x5-reuse}). We note that the coordinated approaches significantly reduce delays relative to uncoordinated approaches as any observation beyond the short local observation horizon is advantageous. In experiments with non-local observation, we note that increasing the non-local observation length from 5s to 30s for TuSeRACT can reduce delays by as much as 20\%.

\begin{table}
	\centering
	\begin{tabular}{|c|c|c|c|}  
		\toprule
		Approach & \multicolumn{3}{c}{Change in mean delay over SURTRAC (\%)}  \\
		& 4000 vph & 5000 vph & 6000 vph \\
		\midrule
		UTuS10 & -15.1 $\pm$ 6.8 &-11.0 $\pm$ 11.0 &-4.2 $\pm$ 11.7 \\
		UTuS10 + gs & -40.1 $\pm$ 4.8 & -43.3 $\pm$ 3.8 & -42.2 $\pm$ 3.4 \\
		CTuS10 & -42.8 $\pm$ 5.3 & -30.8 $\pm$ 7.2 & -20.7 $\pm$ 5.5 \\
		CTuS10 + gs& -51.6 $\pm$ 2.6 & -48.8 $\pm$ 4.0 & -46.8 $\pm$ 4.6\\
		\bottomrule
	\end{tabular}
	\caption{Effect of guiding CP Optimizer search with an initial solution for a 5x5 grid network.}
	\label{tab:5x5-reuse}
\end{table}

\section{Conclusion}
In this paper, we propose a sampling-based approach to traffic signal control in the presence of vehicle turn-induced uncertainty. We show experimentally that our approach provides significant reductions (of up to 50\%) in delay over SURTRAC, the leading approach to distributed real-time traffic signal control. Initial experiments show that sampling is a promising approach to tackle uncertainty in this domain, and that significant performance improvement can be achieved over an existing approach with very few samples (10), and in reasonable time.

\bibliography{Draft}

\end{document}